\documentclass[showpacs,preprintnumbers,nofootinbib,amsmath,amssymb]{revtex4}

\usepackage[pdftex]{graphicx}
\usepackage{amsmath}
\usepackage{amssymb}
\usepackage{epstopdf}
\usepackage{graphicx}
\usepackage{enumerate}

\usepackage{dcolumn}
\usepackage{bm}

\newcommand{\be}{\begin{equation}}
\newcommand{\ee}{\end{equation}}
\newcommand{\bear}{\begin{eqnarray}}
\newcommand{\eear}{\end{eqnarray}}
\newcommand{\ba}{\begin{array}}
\newcommand{\ea}{\end{array}}

\newskip\humongous \humongous=0pt plus 1000pt minus 1000pt

\newif\ifdtup

\def\oldreffmt#1{\rlap{[#1]} \hbox to 2\parindent{}}

\def\figfmt#1{\rlap{Figure {#1}} \hbox to 1in{}}

\def\beq{\begin{equation}}
\def\eeq{\end{equation}}
\def\bea{\begin{eqnarray}}
\def\eea{\end{eqnarray}}

\def\bq{\begin{quote}}
\def\eq{\end{quote}}

\relax

\newdimen\tdim
\tdim=\unitlength
\def\bar{\overline}

\begin{document}

\preprint{UCI-HEP-TR-2013-12}

\title{The Pitfalls of Dark Crossings}

\author{Stefano Profumo}
\email{profumo@ucsc.edu}\affiliation{Department of Physics, University of California, 1156 High St., Santa Cruz, CA 95064, USA}\affiliation{Santa Cruz Institute for Particle Physics, Santa Cruz, CA 95064, USA} 

\author{William Shepherd}
\email{wshepher@ucsc.edu}\affiliation{Department of Physics, University of California, 1156 High St., Santa Cruz, CA 95064, USA}\affiliation{Santa Cruz Institute for Particle Physics, Santa Cruz, CA 95064, USA} 

\author{Tim M. P. Tait}
\email{ttait@uci.edu}\affiliation{Department of Physics \& Astronomy,
University of California, Irvine, CA 92697} 

\begin{abstract}
We explore the connection between pair production of dark matter particles at collider experiments and annihilation of dark 
matter in the early and late universe, with a focus on the correlation between the two time-reversed processes. We consider 
both a model-independent effective theory framework, where the initial and final states are assumed to not change
under time-reversal, and concrete UV-complete models within the framework of supersymmetric extensions to the Standard 
Model. 
Even within the effective theory framework (where crossing symmetry is in some sense assumed), 
we find that the predictions of that symmetry can vary by orders of magnitude depending on the details of the selected effective 
interaction. Within the supersymmetric models we consider, we find that there is an even wilder variation in the expectations one 
can derive for collider observables based on cross-symmetric processes such as having a thermal relic or given indirect dark 
matter detection rates. We also explore additional ``pitfalls'' where na\"ive crossing symmetry badly fails, including models with 
very light mediators leading to
Sommerfeld enhancements and/or dark matter bound states.
\end{abstract}

\pacs{95.35.+d}

\setcounter{page}{0}
\maketitle

\section{Introduction}
\label{sec:intro}

Understanding the particle nature of dark matter is likely to be a 
decisive portal towards unveiling the broader picture of physics beyond the Standard Model. Extracting 
information on dark matter as a fundamental particle relies on the assumption that such particles interact, 
even if only weakly, with particles in the Standard Model. Processes thus might exist whereby one could 
obtain more or less direct information about such interactions. 
The notion that given a process (for example, dark matter elastic scattering off of quarks contained in nuclei) 
a ``conjugated'' or cross-symmetric process exists at quantifiable levels (for example, dark matter pair 
annihilation into quark pairs in the galaxy or quarks annihilating into pairs of dark matter particles at the LHC) 
is key to the idea of ``complementarity'' in the search for signatures from dark matter \cite{Bauer:2013ihz}. 

The key ways in which we can hope to learn about weak-scale weakly-interacting massive particles, commonly known
as WIMPs, can be classified in terms of general processes of the following type, where $\chi$ indicates a WIMP 
and SM any Standard Model particle:
\begin{itemize}
\item {\em Direct Detection}: $\chi+{\rm SM}\to\chi+{\rm SM}$;
\item {\em Indirect Detection}: $\chi+\chi\to{\rm SM}+{\rm SM}$;
\item {\em Collider Searches}: ${\rm SM}+{\rm SM}\to \chi+\chi$.
\end{itemize}
In addition, processes from the three categories above can lead to other probes of the nature of dark matter 
(for example elastic scattering sets the temperature for kinetic decoupling of dark matter in the early universe, which in 
turn determines the small-scale cutoff to the dark matter power spectrum \cite{Cornell:2013rza}, and the pair annihilation 
processes determine the thermal relic density of WIMPs from the early universe).

Relating the processes listed above in some quantitative way is an important task, and one that has 
historically lead to setting important benchmarks for WIMP searches. For example, the assumption that the 
pair annihilation of WIMPs in the early universe (specifically, at the time of chemical freeze-out) is the same 
as the annihilation rate in the late universe (for example, in the Milky Way today) produced the expectation for 
the natural scale of the relevant thermally averaged pair-annihilation cross section times relative 
velocity $\langle\sigma v\rangle_0\simeq3\times 10^{-26}\ {\rm cm}^3/{\rm s}$. This figure, motivated by the 
requirement of the thermal WIMP relic density from the early Universe matching the observed average dark 
matter density, is often considered the benchmark for indirect WIMP searches with gamma rays, high-energy 
neutrinos or antimatter. The weak-interaction level cross section implied by that pair-annihilation cross section 
also motivates collider studies where, for example, one pair-produces dark matter particles with a cross 
section at the level implied by the time-reversed process, plus jets or photons \cite{Beltran:2010ww}. The 
cross-symmetric $\chi+{\rm SM}\to\chi+{\rm SM}$ process, again with a similarly motivated 
cross-section level, is employed to set the scale for direct dark matter searches.

Exploring the validity of crossing symmetry arguments to set the stage for making use of complementarity in the 
search of dark matter is thus a very important task. Clearly, the degree to which crossing symmetry can be used 
depends on the assumed model for the particle sector comprising the dark matter. In principle, however, it should 
be possible to use constraints from one class of processes (for example collider searches) to set bounds on or to 
delineate the relevant parameter space for another class of processes (for example direct detection). This type of 
exercise has been carried out along several model-dependent avenues, e.g. in the context of supersymmetric 
dark matter models [see e.g. \cite{Cahill-Rowley:2013dpa} for a recent study] 
or in the context of universal extra-dimensional models [e.g. \cite{Bertone:2010ww}]. Additionally, the possibility 
that heavy physics described by an effective contact interaction 
mediates the coupling of the dark matter to Standard Model particles has been 
widely explored. This has lead to detailed predictions relating direct detection and collider searches for theories 
featuring higher-dimensional operators representing heavy states that have been integrated out of the effective 
theory \cite{Goodman:2010yf,Bai:2010hh,Goodman:2010ku,Rajaraman:2011wf,Fox:2011pm,Fox:2012ee}.

While crossing symmetry is undoubtedly an important theoretical resource to produce predictions for complementary 
dark matter search channels, it is equally important to appreciate how crossing symmetry arguments might fail. 
Perhaps the most natural place where crossing symmetry is expected to provide accurate guidance 
is in the indirect detection versus 
collider direction, processes that are essentially simply related by time-reversal. In addition, models with the same 
thermal relic abundance can lead to drastically different outcomes for the directly related indirect detection rates as 
well as for collider searches.

In the present study, we explore examples of the ``pitfalls'' of crossing symmetry arguments in the context of 
dark matter searches. We start with the point that in effective theories, the time-reversal crossing symmetry relating 
indirect detection (and the relic density) to collider searches can fail drastically, despite the fact that 
crossing symmetry is, in some sense, assumed by construction (sec.~\ref{sec:effective}). 
We show in sec.~\ref{sec:susy} how UV-complete models of dark matter, such as neutralinos in
supersymmetric models, can produce widely different outcomes for indirect detection and collider searches even 
while giving rise to the correct thermal relic abundance, and for the same lightest neutralino mass. 
Finally, we discuss models where crossing symmetry fails badly, either because the dynamics of the dark sector contains light
mediators, leading to richer phenomena than is expected for heavier mediators, or because of the presence of inelastic processes in dark matter scattering (sec.~\ref{sec:badlyfails}). 
We present our conclusions in sec.~\ref{sec:conclusions}.

\section{Manifest Crossing Symmetry Failing to Manifest}
\label{sec:effective}

The first class of models we consider is the case in which the particles mediating dark matter's interactions
with the Standard Model are very heavy, leading to a description in terms of contact interactions in the context of an
effective field theory (EFT). 
We work in the simplest constructions where the dark matter is a SM singlet, and, to be concrete,
specialize to the case where it is a fermion\footnote{
The dark matter could be Dirac or Majorana, with the
key difference between the two being that certain 
bi-linears (e.g. the vector and tensor), from which contact interactions with Standard Model fields 
can be constructed, identically vanish in the Majorana case.}.  
It is straight-forward to generalize our
analysis to dark matter of arbitrary spin.
We choose the specific EFT implementations of
Ref.~\cite{Goodman:2010yf,Goodman:2010ku}, which consider the lowest-dimension operators
with a particular Lorentz structure, keeping the leading terms as dictated by
minimal flavor violation \cite{D'Ambrosio:2002ex}.

The relation between direct detection and collider signals within these models has been thoroughly considered 
\cite{Goodman:2010yf,Bai:2010hh,Goodman:2010ku,Rajaraman:2011wf,Fox:2011pm,Fox:2012ee,Haisch:2013uaa,Haisch:2012kf,Fox:2012ru,Lin:2013sca,Dreiner:2013vla}.  The picture that emerges
suggests that collider and direct searches enjoy a high degree of complementarity, with colliders particularly
shining for low mass dark matter and for interactions leading to predominantly spin-dependent elastic
scattering, and direct searches providing stronger limits for interactions leading to
spin-independent scattering and for cases where the collider energies
can resolve a light color-singlet 
mediator \cite{Bai:2010hh,Goodman:2011jq,Shoemaker:2011vi,An:2012va,Frandsen:2012rk}.
Some of the most important connections to indirect detection signals have also been explored
\cite{Beltran:2008xg,Cao:2009uw,Goodman:2010qn,Cheung:2010ua,Cheung:2011nt,Cheung:2012gi,Rajaraman:2012db,Frandsen:2012db,Chen:2013gya}, and indicate that indirect detection is also a very important
part of the program of covering the space of allowed interactions. 
Here, our focus will be
primarily on the correlations between the annihilation cross section (with particular interest in inferring the relic density)
and the collider production cross section.

\subsection{Leading Versus Sub-leading Interactions}

In some sense, crossing symmetry is a ``built-in'' component of effective theories where the only 
interactions of dark matter with Standard Model fields is via local operators.  However, a 
given UV complete theory is
likely to populate more than one operator with similarly-sized coefficients, and (as discussed 
in detail below), there
is no guarantee that the same operator will dominate for all three processes.  It is well-known
that the relativistic behavior of dark matter bilinears is most easily organized by the parity
of the bilinear. However, Lorentz and gauge-invariant Lagrangians are naturally built out of Weyl spinors
leading to interactions organized in terms of left or right-chiral projections containing both parity-even
and parity-odd terms.  Thus, it is likely that a term which is leading for direct detection will be accompanied
by a related term which will lead for non-relativistic annihilation.  Comparing annihilation rates and colliders
is more subtle, because the mono-object signal produces dark matter particles relativistically and thus
produces dark matter somewhat agnostically with respect to the underlying Lorentz structure.  In a typical
case, one could easily expect that a single operator will dominate annihilation, but comes with 
chiral-related terms that contribute at colliders, leading to an order one mismatch between the two descriptions
even when the EFT is a good description.

\subsection{Suppressed Annihilations}

We note that at tree level, annihilations of any Majorana fermion dark matter into Standard Model fermions are
suppressed. This is a well-known result, and commonly referred to as $p$-wave suppression.
It is worth mentioning, however, that the annihilation is only strictly $p$-wave for
interactions which are scalar or axial-vector in 
nature\footnote{The $p$-wave nature of the annihilations of fermions through scalar and axial-vector 
operators is simple enough to understand intuitively; left- and right-handed fields are indistinguishable 
at zero velocity, so something which couples to those states with opposite strength will give no 
contribution in that limit. Scalars and axial vectors do precisely this.}. 
In particular, a pseudoscalar-type interaction is allowed, and is suppressed by SM chirality, yet 
gives rise to $s$-wave annihilations. 
This distinction may seem to be purely academic, but when considering the predictions for indirect 
dark matter searches versus those for collider production the difference in mechanism of the 
suppression becomes very important. Chirality suppression is energy-independent, but it does depend on the mass 
of the SM fermion involved in the process. This means that dark matter sufficiently heavy to annihilate into 
top quarks is only very weakly suppressed, while attempting to produce dark matter from the light 
quarks which are prevalent in the proton is very highly suppressed. $p$-wave suppression, on the other 
hand, is directly connected to the low velocity of annihilating WIMPs in the halo, a concern which is totally 
alleviated when WIMP pairs are produced relativistically at the LHC. 
If we consider instead a Dirac dark matter candidate, two new types of interactions are allowed. 
Tensor-type interactions are also suppressed by SM chirality, but are once again 
allowed to proceed in the $s$-wave, while vector-type interactions are not suppressed by 
anything but the scale of the interaction itself.

Even once we `know' the dark matter spin and remaining within the simple framework of 
the EFT, what we assume about an observed signal has significant effects on our 
extrapolation to other experiments. Depending on the Lorentz structure of  the operator, dark matter 
annihilation processes can be unsuppressed, $p$-wave suppressed, suppressed by quark masses in 
accordance with the conjecture of MFV, or suppressed by both of the above. 
Operators which are suppressed by SM chirality will cause dark matter to annihilate more 
readily than be produced at colliders (assuming it is heavy enough to annihilate into 
quarks which are rare in the proton), while operators which allow only $p$-wave 
interactions are more easily probed at colliders than in annihilations. When both of these 
suppressions are present, the $p$-wave suppression is generally more important for the 
suppression scales which are accessible to experiment (and which can lead to an appropriate 
relic density), so that collider searches would be the more promising search, but neither search 
has much reach. We have listed the most commonly considered operators in table \ref{tab:ops} and 
indicated which suppressions, if any, are present for those operators. 
The table reveals that this space of operators covers all combinations of suppression, ranging from the
vector operator which is completely unsuppressed to the scalar interaction which is both
$p$-wave {\em and} chirally-suppressed.

\begin{table}
\caption{\label{tab:ops}Common operators and their suppression mechanisms with regard to
non-relativistic annihilation.}
\begin{tabular}{|c|c|c|}
\hline
~~~Operator&Suppressed by Chirality~~~& ~~~$p$-wave Suppressed~~~\\
\hline
$\bar\chi\chi~\bar qq$&Yes&Yes\\
$\bar\chi\gamma^5\chi~\bar q\gamma^5q$&Yes&No\\
$\bar\chi\gamma^\mu\chi~\bar q\gamma_\mu q$&No&No\\
$\bar\chi\gamma^\mu\gamma^5\chi~\bar q\gamma_\mu\gamma^5q$&No&Yes\\
$\bar\chi\sigma^{\mu\nu}\chi~\bar q\sigma_{\mu\nu}q$&Yes&No\\
\hline
\end{tabular}
\end{table}

As a particular example, it is instructive to compare the pseudo-scalar and axial vector operators:
\bear
 \bar\chi\gamma_5 \chi ~ \bar q\gamma_5 q ~~~~~~ {\rm versus} ~~~~~
 \bar\chi\gamma^\mu\gamma_5\chi ~ \bar q\gamma_\mu\gamma_5 q.
 \eear 
 The first is quark-mass suppressed and thus difficult to see at the LHC, while the second is $p$-wave 
 suppressed in annihilations. If we had e.g. some signal from collider searches of missing energy production, 
 we could attempt to extrapolate using either of these operators to predict what cross section
 should be seen in indirect detection.
 On the one hand, the mass-suppressed operator results in a difference of sea quark versus valence quark
 parton distribution functions in matching the collider signal, leading to an increase in the necessary
 interaction strength to produce a given collider result relative to an unsuppressed operator, and ultimately
 a larger annihilation cross section into heavy quarks.  On the other
 hand, the $p$-wave operator picks up a large $v^2$ suppression for annihilations in the galactic halo.
As a result, in this case we estimate that the translation of collider observation to annihilation cross section would differ
by more than eight orders of magnitude between these two operators.  Of course, ultimately this haziness
can be understood as a blessing -- if one expected to be sensitive in an indirect search, failing to
observe the signal would strongly suggest the axial-vector operator as the origin of the collider
observation.  Further information could be provided by direct detection experiments, since the
pseudo-scalar operator is velocity-suppressed in direct detection, whereas the axial-vector interaction
would lead to a spin-dependent signal.

\section{Tricky Supersymmetric Thermal Relics}\label{sec:susy}

Crossing symmetry can also be subtle in UV complete models of dark matter.  Here we focus on the
concrete case of the Minimal Supersymmetric Standard Model (MSSM). 
We intend to provide examples where:
\begin{enumerate}[(i)]
\item the guideline of having a thermal relic density matching the observed dark matter density 
fails at making unique predictions for indirect dark matter detection rates, and
\item where predictions for collider rates are wildly different, even if the dark matter is a 
good thermal relic, and even if dark matter indirect detection rates  are comparable and the dark matter particle mass is identical.
\end{enumerate}
All of the models we select have a neutralino as the lightest supersymmetric particle, 
and feature a thermal relic neutralino density which saturates the observed cosmological 
dark matter density today. We select points which have particular indirect dark matter detection  
phenomenologies connected with the various possible dominant mechanisms of dark matter 
annihilation, in the early and in the late universe.  Our points are chosen to
illustrate certain cases; for examples of 
more complete exploration of dark matter in the MSSM, see \cite{Cahill-Rowley:2013dpa,Cotta:2011ht,Cotta:2011pm,CahillRowley:2012cb}.

For all models, we take the ratio of the neutral Higgses vacuum expectation values, 
$\tan\beta=15$, and, unless otherwise specified, we take 
$m_A=1$ TeV, $M_2=M_3=1$ TeV, $M_1=200$ GeV, all tri-scalar couplings are set to zero and, again 
unless otherwise specified (for Model 3), we take the scalar soft supersymmetry-breaking sfermion 
masses to be 8 TeV. The model parameters quoted in Tab.~\ref{tab:benchmarks} are all tuned to 
produce the observed dark matter abundance in the form of thermal relic neutralinos. In addition, 
in all models the Higgs mass is compatible with recent results from the LHC collaborations within the experimental and theoretical uncertainties.
Tab.~\ref{tab:benchmarks} collects the relevant input parameters, as well as the output lightest 
neutralino mass $m_{\chi^0_1}$ and the lightest CP-even neutral Higgs $m_h$. All masses and 
mass parameters are given in GeV in the table.

\begin{table}
\begin{tabular}{|c|c|c|c|c|c|c|}
\hline \\[-0.3cm]
~~Model~~ & ~~$M_1$~~ & ~~$\mu$~~  & ~~$m_{\tilde\tau_1}$~~ & ~~$m_A$~~ & 
~~$m_{\chi^0_1}$~~ & ~~$m_h$~~ \\[0.1cm]
\hline \\[-0.3cm]
Model 1 & 200 & 239 & $8000$ & $1000$ & 182 & 127 \\[0.1cm]
Model 2 & 194 & 220 & $8000$ & $1000$ & 172 & 128.6 \\[0.1cm]
Model 3 & 200 & $1000$ & 201 & $1000$ & 199 & 130 \\[0.1cm]
Model 4 & 200 & $1000$ & $8000$ & 454 & 199 & 127 \\[0.1cm]
Model 5 & 200 & $1000$ & $8000$ & 391.5 & 199 & 125 \\[0.1cm]
\hline
\end{tabular}
\caption{Selected input and output parameters (in GeV)
for the five supersymmetric benchmark models; see the text for additional details.}\label{tab:benchmarks}
\end{table}

In Models 1 and 2, the thermal relic density is set by the lightest neutralino having a sizable 
higgsino fraction, and by coannihilation with the lightest chargino and the next-to-lightest neutralino. 
This is achieved by tuning the higgsino mass parameter $\mu$ to values close to the bino soft 
supersymmetry breaking mass term $M_1$. The difference between our ``benchmark'' Model 1 and 
Model 2 is that in the latter case the lightest neutralino mass was tuned to be slightly below the top 
quark mass threshold. In the early universe, and specifically at thermal freeze-out, when 
$T\simeq m_{\chi^0_1}/20\approx 8.5$ GeV, pair-annihilation into top quarks was kinematically 
accessible for large enough neutralino velocities, while in the late universe the top annihilation channel is 
entirely closed. This provides an example of how the relevant pair annihilation cross section in the 
early versus late universe can differ. Note that for our particular parameter space choice, the effect 
is not particularly dramatic, namely reducing the pair annihilation cross section today by approximately 
a factor of 2 with respect to the early universe (see Tab.~\ref{tab:indirdet}). This is due to the fact that for 
both Models 1 and 2 the pair-annihilation into $W^+W^-$ or $ZZ$ final states is relatively important, 
on the same order as the pair-annihilation into top quark pairs.

Model 3 is an instance of a model where the thermal relic density is set by the co-annihilation 
with the lightest stau, whose mass was fine-tuned to be approximately 1 GeV above the lightest 
neutralino mass. In this case, the neutralino pair annihilation cross section is highly suppressed 
(in fact, $\langle\sigma v\rangle\sim 10^{-28} \ {\rm cm}^3/{\rm s}$) as the  effective cross section 
setting the thermal relic density is dominated by stau annihilation (45\%) and co-annihilation (50\%).

Models 4 and 5 feature a Higgs particle with a mass close to twice the lightest neutralino mass, 
thus opening a resonant pair-annihilation $s$-channel where the lightest neutralino annihilates via a 
close-to-on-shell scalar into (typically) fermion-antifermion final states. In Model 4, the lightest 
neutralino mass is below the $m_A/2$ threshold, while for Model 5 we tune $m_A$ so that $m_\chi>m_A/2$. In both cases the late-time cross section is off-resonance. However, due to finite-temperature effects, the late-time annihilation rate is larger than the naive expectation (from $s$-wave annihilation) for Model 5, while it's smaller for Model 4, as we explain below.

 The late-universe 
(i.e. $T=0$) pair-annihilation cross section for Model 4 is {\em smaller} than the reference thermal 
cross section ($3\times 10^{-26}$ cm$^{-3}$/s) at a value of 6.4$\times 10^{-27}$ cm$^{-3}$/s: the 
early-universe finite-temperature cross section, responsible for setting the thermal relic density, results from the convolution of a temperature kernel times the cross section, which at $\sqrt{s}>2m_\chi$ is increasingly more ``on-resonance''. Vice versa, for 
Model 5 the resonance is increasingly off shell in the early universe, while it's closer to being on shell in the late 
universe, resulting in a pair-annihilation cross 
section today larger than the naive thermal $s$-wave prediction, at 5.2$\times 10^{-26}$ cm$^{-3}$/s.
 
In what follows we outline predictions for collider searches (sec.~\ref{sec:coll}) and for 
indirect detection searches (sec.~\ref{sec:indirect}) for the five benchmark models outlined above, 
and show that even within the restrictive framework of the MSSM, models with the same 
neutralino mass and with the correct thermal neutralino relic abundance produce a very 
broad set of outcomes for ``cross-symmetric" processes.

\subsection{Collider Predictions}\label{sec:coll}

For each of the benchmark models described above, we calculate the pair production 
cross section for supersymmetric particles. The total production cross sections as well as a 
breakdown into subprocesses of particular interest, including direct production of pairs of LSPs
($\chi^0_1$), the sum of all pairs of neutralinos and/or charginos (Inclusive $\chi \chi$),
and pairs of lightest staus ($\widetilde{\tau}_1^+ \widetilde{\tau}_1^-$)
are shown in Table \ref{tab:sigma}. Note that outside of Model 3, the stau pair production cross section vanishes since $m_{\widetilde{\tau}_1}$ is set to 8 TeV, as indicated in table \ref{tab:benchmarks}. The heavier $\widetilde{\tau}$ is always chosen to have a mass of 8 TeV. As all of the 
colored particles have been chosen to be 
heavy enough to be decoupled, electroweak production mechanisms are 
dominant. The usual jets and missing energy search corresponds to pair production of the lightest 
neutralino with an additional hard initial state radiation (ISR) jet, and therefore is bounded from 
above by the $\chi^0_1$ pair production cross section. Note that in all of the models we consider 
here, this cross section is smaller than a femtobarn.

\begin{table}
\begin{tabular}{|c|c|c|c|c|}
\hline \\[-0.3cm]
~~~Model~~~ & ~Total EW $\sigma$ (fb)~ & ~$\chi^0_1 \chi^0_1$ $\sigma$ (fb)~ & 
~Inclusive $\chi \chi$ $\sigma$ (fb)~ & ~$\widetilde{\tau}^+ \widetilde{\tau}^-$ $\sigma$ (fb)~ \\[0.1cm]
\hline \\[-0.3cm]
Model 1 & 625 & 0.2 & 625 & 0 \\[0.1cm]
Model 2 & 850 & 0.3 & 850 & 0 \\[0.1cm]
Model 3 & 9.9 & 0.001 & 2.8 & 7.1 \\[0.1cm]
Model 4 & 2.8 & 0.001 & 2.8 & 0 \\[0.1cm]
Model 5 & 2.8 & 0.001 & 2.8 & 0 \\[0.1cm]
\hline
\end{tabular}
\caption{\label{tab:sigma}LHC pair-production cross sections for the stated electroweakly-charged particles in each of the five benchmark models}
\end{table}

Even in our ``vanilla'' benchmark Model 1, where the neutralino is well-tempered and
mixing is the factor which drives the 
relic density, direct dark matter production at the LHC is very small. While these models do 
respect crossing symmetry well, by construction, annihilations do not proceed dominantly to quark 
final states, and thus a production cross section comparable to the annihilation cross section 
cannot be achieved by a hadron collider. This illustrates a first crossing symmetry issue: collider 
searches are sensitive to having very selected final states in the cross-symmetric annihilation 
process, and such final states are not generic. A ``weak boson collider'' would best probe the 
physics which leads to the dark matter relic density in these situations. Model 2, featuring the 
sub-top quark mass threshold, also exhibits similar behaviour (since the top threshold does not impact 
the collider production cross section).

We also see strong suppression of SUSY production cross sections when the dominant contributor to the 
effective pair-annihilation cross section in the early universe is coannihilation, in our case 
with staus. This allows the dark matter to be much more weakly coupled to standard matter than 
naively expected from crossing symmetry arguments, while still generating the correct relic density, 
further divorcing the model from the expectation of a comparable production cross section. In this case 
a tau collider, while perhaps unlikely to be realized in the foreseeable future, 
would be the optimal choice to produce dark matter pairs while probing the physics 
which leads to the relic density.

Models 4 and 5 feature an $A$- (and $H$-)
funnel to get the correct relic density of dark matter, and thus have very little production cross section of 
any SUSY particles at all at the LHC, as those states couple very weakly to light quarks. In this case, the 
best probe of dark matter phenomena is contained within Higgs physics, where the heavier state can be 
produced and its couplings measured. However, even with the discovery of new heavy scalars, 
additional complications can arise. As discussed above, the distinction between Models 4 and 5 is that 
in one case the dark matter is just lighter than $M_A/2$, and in the other just heavier. In the second 
case the invisible width will actually vanish, and no notable sign of dark matter will be present in 
on-shell Higgs processes. It is only with an independent dark matter mass measurement that the 
funnel origin of the relic abundance in this model can be divined.

\subsection{Indirect Detection Predictions}
\label{sec:indirect}

\begin{table}
\begin{tabular}{|c|c|c|c|c|c|c|}
\hline \\[-0.3cm]
~~~Model~~~ & ~$\langle\sigma v\rangle_0$~ & 
~~$\langle\sigma v\rangle_{\gamma\gamma}/\langle\sigma v\rangle_0$~~ & 
~~$\phi_\gamma(E>1\ {\rm GeV})$~~ & ~~$\phi_{\gamma\gamma}$~~ & 
~~$\sigma_{\chi p}^{\rm SI}$~~ & ~~$\sigma_{\chi p}^{\rm SD}$~~\\[0.1cm]
\hline \\[-0.3cm]
Model 1 & $2.03\times 10^{-26}$  & 3.50$\times 10^{-5}$ & 1.73$\times 10^{-9}$ & 6.50$\times 10^{-15}$  &  1.62$\times 10^{-8}$ & 1.53$\times 10^{-4}$  \\[0.1cm]
Model 2 &  $1.37\times 10^{-26}$ & 1.32$\times 10^{-4}$ &1.26$\times 10^{-9}$ &  2.31$\times 10^{-14}$ &  1.81$\times 10^{-8}$ &  2.01$\times 10^{-4}$ \\[0.1cm]
Model 3 & $9.20\times 10^{-29}$ & 8.99$\times 10^{-3}$ & 1.91$\times 10^{-12}$ &  4.44$\times 10^{-15}$ &  5.38$\times 10^{-11}$ & 1.34$\times 10^{-7}$  \\[0.1cm]
Model 4 &  $6.39\times 10^{-27}$ &1.55$\times 10^{-8}$& 2.92$\times 10^{-10}$ & 5.69$\times 10^{-19}$  & 2.50$\times 10^{-10}$  &  1.34$\times 10^{-7}$ \\[0.1cm]
Model 5 & $5.18\times 10^{-26}$ & 9.19$\times 10^{-7}$ &2.86$\times 10^{-9}$ & 2.73$\times 10^{-16}$ &3.97$\times 10^{-10}$&1.23$\times 10^{-7}$\\[0.1cm]
\hline
\end{tabular}
\caption{Dark matter detection results for the five supersymmetric benchmark models; the first column indicates 
the zero-temperature thermally averaged pair-annihilation cross section times velocity, in units of cm${}^3$/s; the 
second column gives the branching ratio for annihilation into two photons; the third column indicates the 
integrated photon flux above 1 GeV from the direction of the Galactic center, in cm${}^{-2}$s${}^{-1}$, and the 
fourth column the $\gamma\gamma$ flux, in the same units. Lastly, columns 6 and 7 indicate the neutralino-
proton scattering cross section, respectively spin-independent and spin-dependent, in units of pb.}\label{tab:indirdet}\label{tab:indir1}
\end{table}

\begin{figure*}[!h]
\mbox{ \includegraphics[width=0.7\linewidth]{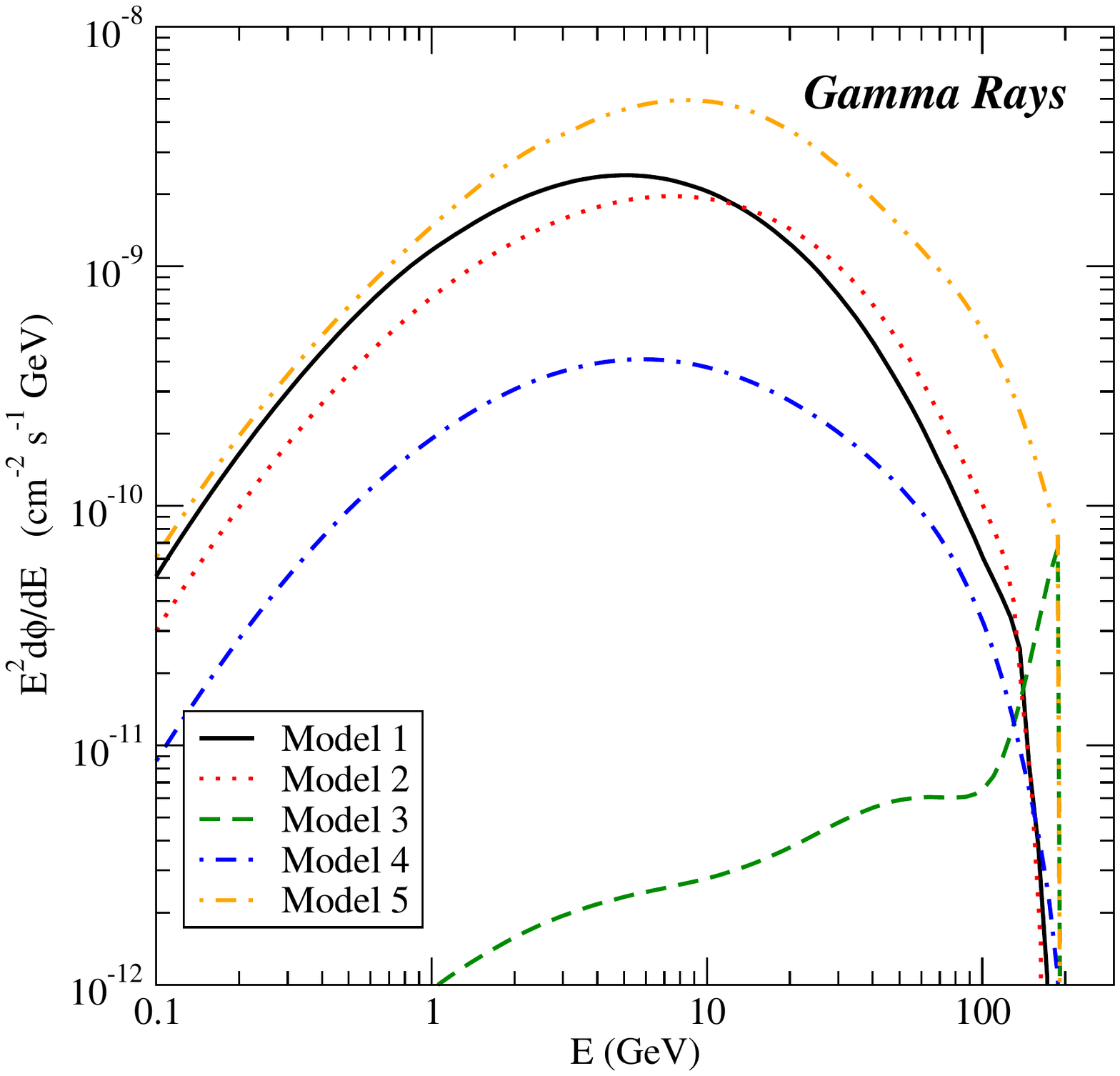} \hspace{-3.cm}  \includegraphics[width=.7\linewidth]{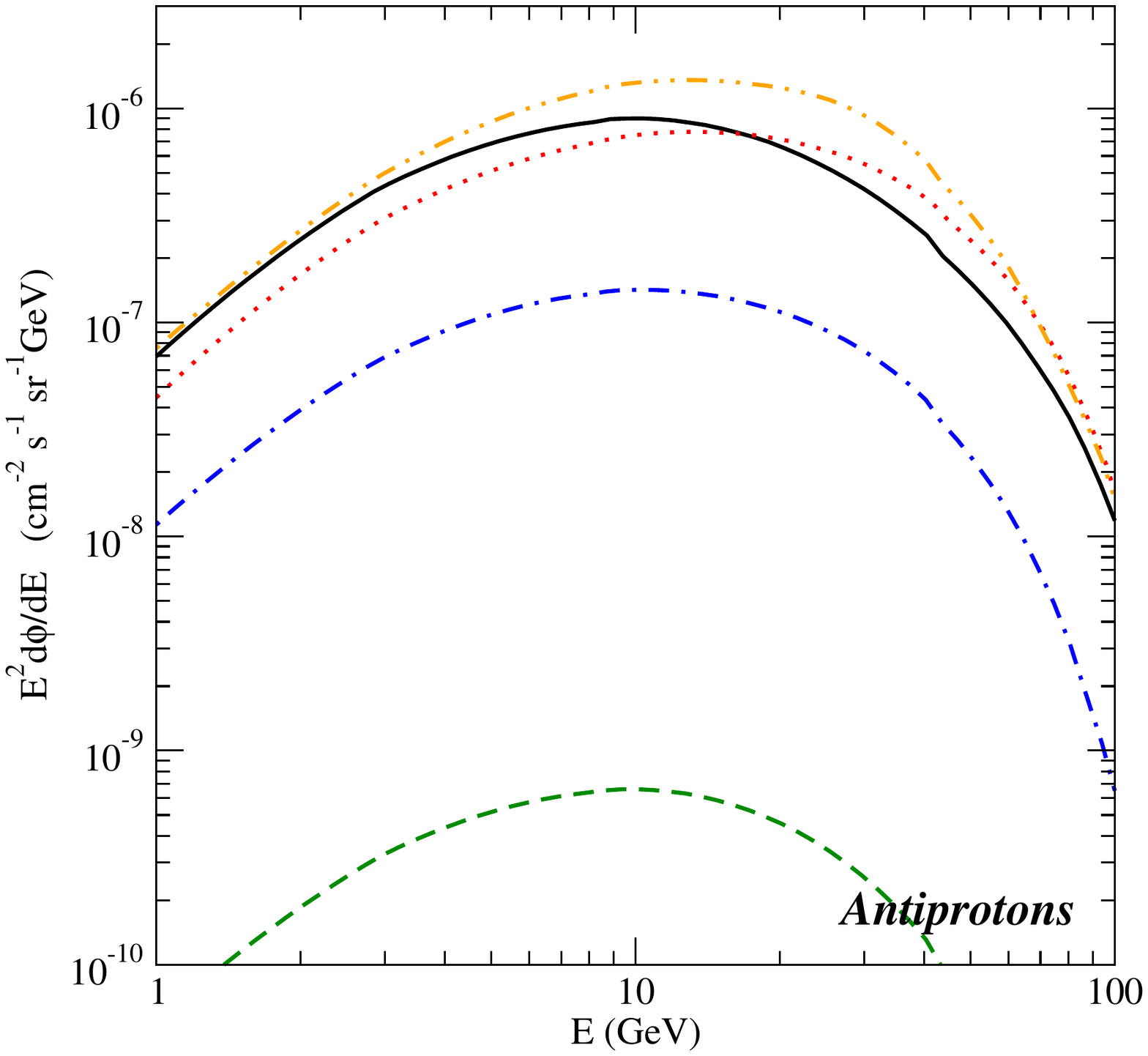}}\\[-0.7cm]
\mbox{ \includegraphics[width=.7\linewidth]{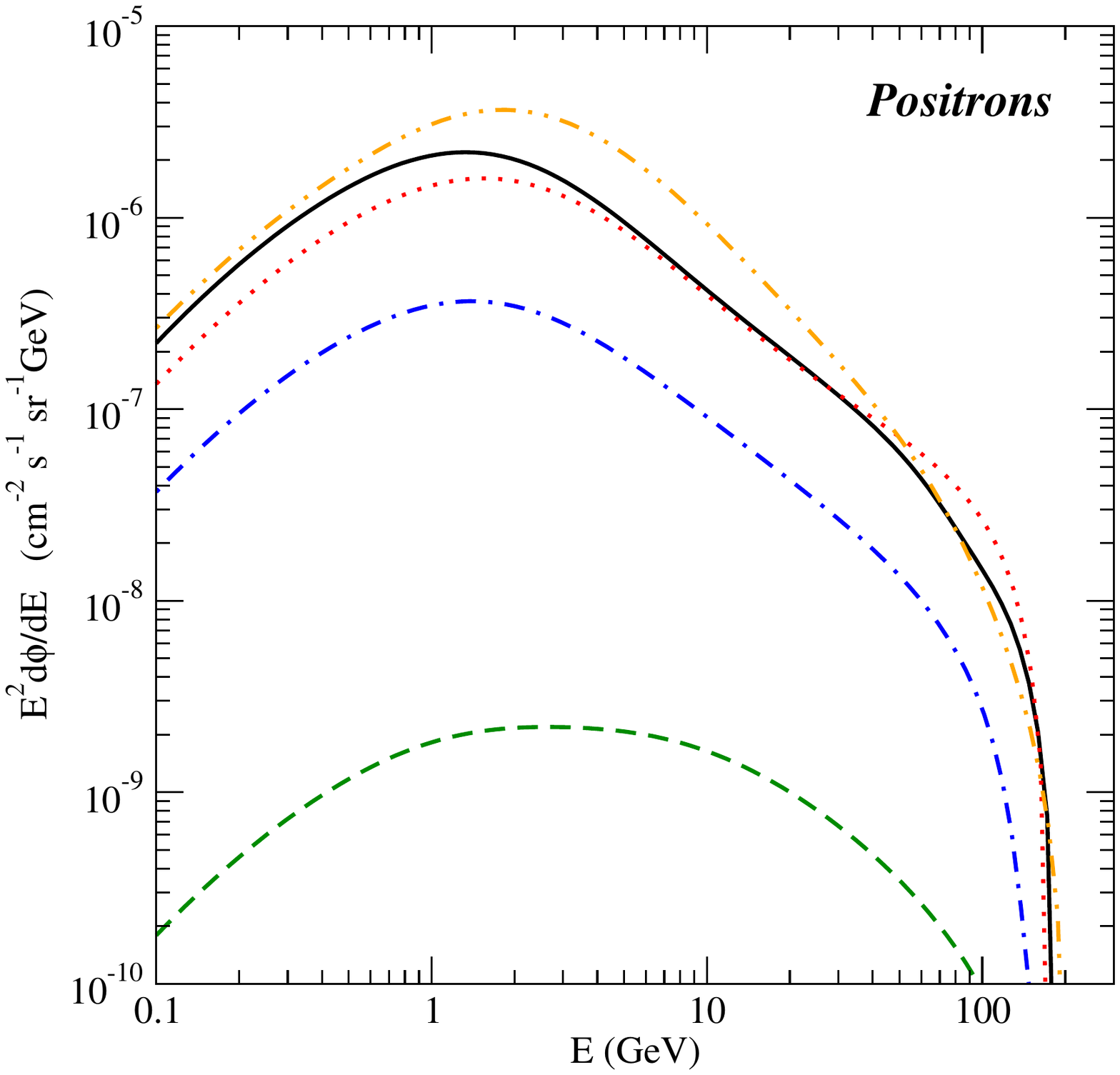} \hspace{-3cm}  \includegraphics[width=.7\linewidth]{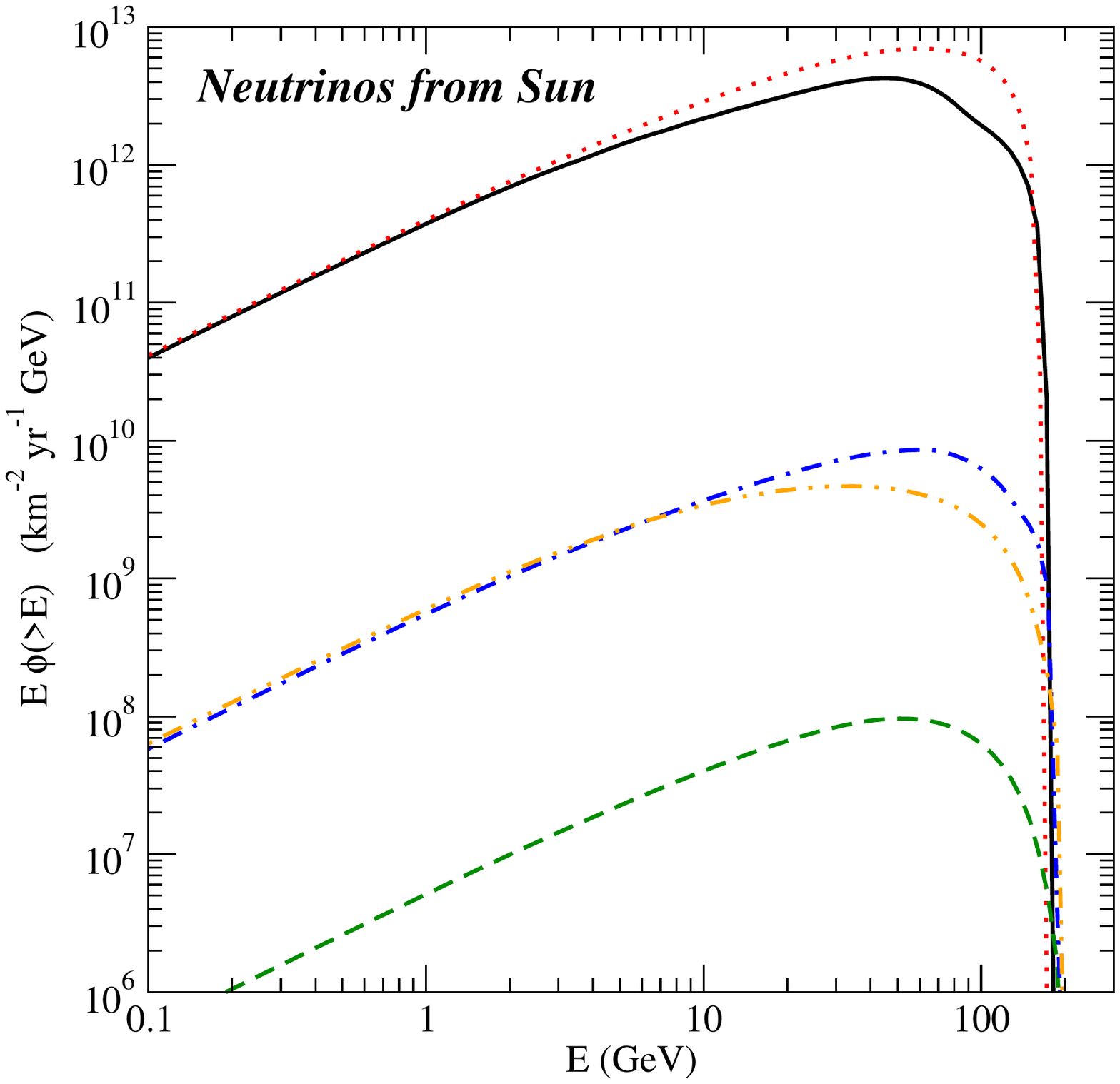}}
  \caption{Differential particle spectra times energy squared for our five ``tricky thermal neutralinos'' benchmark models, for 
gamma rays (top left panel), antiprotons (top right), positrons (bottom left) and neutrinos from the Sun (bottom right). Particle 
energies are in GeV, while spectra are in units of GeV${}^{-}$ cm${}^{-2}$ s${}^{-1}$ sr${}^{-1}$. sr${}^{-1}$. with the exception 
of neutrinos, for which we use GeV${}^{-}$ km${}^{-2}$ yr${}^{-1}$.}\label{fig:spectra}
\end{figure*}

In this section we compare predictions for indirect detection rates for the 5 benchmark ``tricky'' thermal neutralinos 
described above. We collect the relevant detection rates in tables \ref{tab:indir1} and \ref{tab:indir2}. In addition to the 
thermally averaged pair-annihilation cross section, listed in the second column, table \ref{tab:indir1} lists the branching 
ratio into the two-photon final state in column 3. Interestingly, the largest branching ratio corresponds to Model 3, i.e. the 
model with very light staus where stau coannihilation drives the thermal relic density. In this case, in fact, the relevant 
loop process includes the light staus, and, as can be seen from the second column, the total neutralino pair-annihilation 
cross section is quite suppressed, resulting in a relatively large (almost per-cent level) 
$\langle\sigma v\rangle_{\gamma\gamma}/\langle\sigma v\rangle_0$ branching ratio.
We show relevant differential spectra (times particle energy squared) in Figure~\ref{fig:spectra}: the top-left panel shows our 
predictions for gamma rays, the top right for antiprotons, the bottom left for positrons and, finally, the bottom right for neutrinos 
from the Sun. The difference between Models 4 and 5 are ascribed to the different $T=0$ pair-annihilation cross sections, as discussed above. The one exception is the flux of neutrinos from the Sun, which is dominated by the capture rate (similar for the two models) rather than by the annihilation rate, due to capture-annihilation equilibrium.

We calculate the integrated gamma-ray flux from from within a 10$^{-3}$ sr solid angle of
the direction of the Galactic center and for energies above 1 GeV in column 4. We assume a smooth Navarro-Frenk-White
dark matter density profile \cite{Navarro:1996gj}, with a local halo density of $0.3$ GeV/cm$^3$, 
a heliocentric distance of 8 kpc, and a scale radius of 20 kpc. 
Our results are relatively homogeneous for Models 1, 2, 4 and 5, all at the level of $10^{-9}$ cm${}^{-2}$s${}^{-1}$, 
with mild differences due to slightly different $\gamma$-ray spectra (see the top-left panel of Figure~\ref{fig:spectra}) 
and to the slightly differing pair-annihilation cross sections (column 2). The markedly different flux prediction, 
and the strikingly different spectrum shown in the figure, are due to the dominant $\tau^+\tau^-$ final state, 
dictated by the Model's particle spectrum, which features light staus as the only relevant
mediators for the annihilation process. 

Column 5 lists, for the same astrophysical setup as column 4, the predicted monochromatic $\gamma\gamma$ flux. 
In this case we uniformly get suppressed rates, but with larger differences, due to different lightest neutralino compositions 
for Models 1 and 2 and to the different Higgs boson masses for Models 4 and 5, with Model 3 featuring a relatively large 
flux (especially compared to the continuum flux of column 4) due to light staus
running in the loop diagrams.

The last two columns indicate the direct detection cross sections for elastic neutralino-proton interactions, for 
spin-independent (col. 6) and spin-dependent (col. 7) processes. We note the large differences among the 5 
benchmark models for both $\sigma_{\chi p}^{\rm SI}$ and $\sigma_{\chi p}^{\rm SD}$. These are due to 
markedly different higgsino fractions, which are relatively large for Models 1 and 2 and very small for 
models 3, 4 and 5. The relatively small differences between models 4 and 5 are due to the different heavy 
Higgs boson masses. This set of models thus also stands as an example of how models with the same 
neutralino mass and comparable pair annihilation cross sections can have drastically different 
(here at the level of 3 orders of magnitude) direct detection cross sections.

\begin{table}
\begin{tabular}{|c|c|c|c|c|c|c|c|}
\hline
Model & $\phi_\mu^\oplus(E>1\ {\rm GeV})$&$\phi_\mu^\odot(E>1\ {\rm GeV})$ &$\phi_\mu^\oplus(E>100\ {\rm GeV})$&$\phi_\mu^\odot(E>100\ {\rm GeV})$ &$\phi_{\bar p}$&$\phi_{e^+}$&$\phi_{\bar D}$\\
\hline
Model 1 &  1.023$\times10^{-3}$& 462. & 5.31$\times10^{-5}$  & 17.7   & 8.98$\times10^{-9}$ & 2.11$\times10^{-6}$ & 1.61$\times10^{-12}$\\
Model 2 &1.88$\times10^{-3}$ & 896. & 1.09$\times10^{-4}$ & 41.9 &  7.49$\times10^{-9}$ & 1.47$\times10^{-6}$ & 6.21$\times10^{-13}$\\
Model 3 &  2.93$\times10^{-11}$ & 0.013 & 9.97$\times10^{-13}$ & 5.68$\times10^{-4}$ &  6.60$\times10^{-12}$ &  1.82$\times10^{-9}$ & 1.42$\times10^{-15}$ \\
Model 4 & 3.35$\times10^{-8}$ & 1.11 & 6.85$\times10^{-10}$ & 0.070  & 1.42$\times10^{-9}$  & 3.48$\times10^{-7}$ & 2.02$\times10^{-13}$\\
Model 5 & 2.91$\times10^{-7}$ &0.553 &5.88$\times10^{-9}$ &0.0221 &1.32$\times10^{-8}$ & 3.07$\times10^{-6}$ &1.29$\times10^{-12}$\\
\hline
\end{tabular}
\caption{More Indirect detection results for the five supersymmetric benchmark models: columns 2 and 3 indicate the 
integrated muon flux above 1 GeV from the Earth and from the Sun, respectively, in units of km${}^{-2}$yr${}^{-1}$; 
columns 4 and 5 show the same quantity, but integrated above 100 GeV. The last three columns indicate the 
differential antiproton flux at 10 GeV (col. 6), positron flux at 1 GeV (col. 7) and antideuteron flux at 1 GeV (col. 8) 
all in units of GeV${}^{-1}$ cm${}^{-2}$ s${}^{-1}$ sr${}^{-1}$.}\label{tab:indir2}
\end{table}

We continue our survey of indirect detection rates in table \ref{tab:indir2}, which lists the flux of muons from the Earth 
or the Sun produced by neutrinos generated by dark matter annihilation. We show the integrated flux above 1 GeV 
(coloumns 2 and 3) since experimental limits are often given utilizing this threshold, and the physically more relevant 
100 GeV threshold (columns 4 and 5), which gives a more realistic idea of the expected event rate in a telescope 
similar to, for example, IceCube. Fluxes are quoted in units of  km${}^{-2}$yr${}^{-1}$.

The driver for the flux of muons from the Earth and from the Sun is the capture rate, which, in turn, depends on the 
scattering cross section of neutralinos off of nuclei. It is therefore not surprising that we find a hierarchy between 
Models 1 and 2 versus models 4 and 5 that reproduces the hierarchy observed in the last two columns of 
table \ref{tab:indir1}. The slightly softer high-energy neutrino spectrum for Model 5 (see fig.~\ref{fig:spectra}) 
explains the additional suppression for the $>100$ GeV flux for Model 5 versus Model 4. In the case of Model 3, 
equilibration between capture and annihilation does not occur due to the low pair-annihilation cross section, 
and the rates are further suppressed.

The last three columns of table  \ref{tab:indir2} indicate the differential antiproton flux at 10 GeV (col. 6), 
positron flux at 1 GeV (col. 7) and antideuteron flux at 1 GeV (col. 8) all in units of 
GeV${}^{-1}$ cm${}^{-2}$ s${}^{-1}$ sr${}^{-1}$. The rates are generically consistent with the 
pair-annihilation cross section levels quoted in table \ref{tab:indir1}. The spectra we find are 
relatively comparable, with harder spectra found for Model 2 and especially 4, due to the relatively 
large pair-annihilation rate into $\tau^+\tau^-$ pairs (which is suppressed compared to the $b\bar b$ 
final states by factors of the mass ratio squared and of color). As a side note, we find that the positron spectra are in all cases too soft to 
reproduce the positron fraction anomaly reported by AMS-02 \cite{ams02}.

\section{Catastrophic Failures of Crossing Symmetry}\label{sec:badlyfails}

In this section we provide two examples of dark matter models that have received considerable attention in the 
recent past, and where arguments based on crossing symmetry to predict indirect detection and collider rates 
badly fail: models with a so-called Sommerfeld effect (sec.~\ref{sec:somm}) and inelastic dark matter models 
(sec.~\ref{sec:inelastic}).

\subsection{Very Light Mediators and Sommerfeld-like Enhancement}\label{sec:somm}

In a model where dark matter can self-interact by the exchange of very light mediators, 
the non-relativistic cross section
can receive a large Sommerfeld-like enhancement for dark matter scattering at low relative velocity.  
This already implies
a large velocity dependence of the cross section leading to a much larger annihilation rate for 
WIMPs in the Galaxy
(whose velocity dispersion is thought to be order $10^{-3}$) compared to freeze-out, when $v \sim 10^{-1}$.

The mapping between annihilation and colliders may also be
convoluted by the presence of the light mediator.  At colliders,
dark matter is only visible when produced relativistically, and a mediator leading to self-scattering is largely 
irrelevant for the typical missing energy signals.  Furthermore, annihilation may have a large component into
the mediator particles themselves, and this rate is generically difficult to infer from the usual collider
searches for dark matter.  While it may be possible to produce the dark mediator at colliders (for example,
radiated from a final state WIMP in a process which otherwise produces dark matter), the signature of
the mediator is very model-dependent, with some mediators decaying very quickly to SM states
(see e.g. \cite{ArkaniHamed:2008qp} for a discussion of a specific model), and others
stable on collider scales and thus at most appearing as a slight discrepancy in the
distribution of missing momentum \cite{Giudice:2011ib}.

We note that large Sommerfeld-like enhancements are correlated with the presence of
light mediators, for which colliders may see non-negligible effects from
dark matter bound states, or WIMPonia \cite{Shepherd:2009sa}.  Such particles appear as resonances
in SM scattering, and thus imply a new type of signal which can be associated with the presence of the
enhanced annihilation cross section.  The bound state spectra and effective coupling to the SM are
very sensitive to the masses of both the dark matter and the mediator, as well as the dark matter-mediator
coupling.  Detailed measurements of both the bound states and the coupling of unbound WIMPs to SM
fields could potentially be combined to infer the presence of Sommerfeld-like enhancements in
annihilation.

\subsection{Inelastic Dark Matter}\label{sec:inelastic}

Inelasticity completely removes the crossing symmetry between collider production and annihilation, 
since the collider can produce one dark matter particle and one excited state, while annihilation 
will have to proceed through a $t$-channel excited state. 
Dedicated studies of collider sensitivity to dark matter inelasticity have been 
performed \cite{Bai:2011jg}, but they find sensitivity only for splittings much larger than those 
required for the inelastic scattering phenomenon in direct detection which originally motivated these models. 
For smaller splittings the excited dark matter state is effectively
collider stable, and thus inelastic models cannot 
generally be distinguished from ordinary dark matter at colliders.

The long lifetime of the excited state often leads to the dark matter relic density being set, not by 
WIMP-WIMP annihilations, but rather by WIMP-excited state coannihilations, so that the crossing 
symmetry between colliders and annihilations setting the relic density may actually remain largely intact. 
However, at late cosmological times the excited states in many models will have largely decayed 
down to WIMPs and soft photons or neutrinos, such that the annihilations we are currently searching 
for are no longer the same as those which set the relic density.

In the case of one particular model with excited dark matter states the symmetry is 
broken even more completely. For Exciting Dark Matter \cite{Finkbeiner:2007kk} indirect detection has 
already seen indications of dark matter's physics in the INTEGRAL 511 keV excess, but this is not
in fact an 
indication of annihilation at all. Rather, the excess is due entirely to the decays of the excited states 
back down to the WIMP ground state. Depending on parameter choices the excited state could still be 
collider stable or it could decay promptly to a soft electron-positron pair and a WIMP. Distinguishing 
between these two cases would be challenging at the LHC, where soft leptons are useless as
triggers and difficult to identify.

The final member of the class of inelastically scattering dark matter models scatter 
exothermically \cite{Graham:2010ca}.
Such models have recently been invoked \cite{Frandsen:2013cna} to
reconcile a dark matter interpretation of the excess CDMS silicon events \cite{Agnese:2013rvf} 
with the stringent bounds from Xenon experiments \cite{Aprile:2012nq}
(for alternative explanations, see \cite{Cotta:2013jna,Feng:2013vod}).
In this case it is the excited dark state which is the initial state in observable scattering events, 
which necessitates that the excited state have a cosmological lifetime, something not required in other 
inelastic models. With this assumption, a large fraction of the galactic halo can be composed of
the excited states. 
Interestingly, in this case the annihilation does look like a crossing of the scattering process, 
as there are plenty of both states of dark matter available to find each other and annihilate. 
Likewise colliders remain insensitive to the splitting between the two dark states and bound the 
interaction strength without regard to the exothermic phenomenology. Thus, it seems that crossing 
symmetry between colliders and annihilations remains robust, while the relation with direct detection 
is the only case which is altered.

Indeed, manifest crossing symmetry as described in section \ref{sec:effective} is present between the indirect and collider searches for dark matter, but we find ourselves with precisely the same potential for
confusion between operators as were described there. 
There are two candidate operators that could generate appreciable exothermic scattering of dark 
matter on nuclei,
\begin{eqnarray}
\sum_q \frac{m_q}{\Lambda^3}
\bar\chi_1\chi_2~\bar qq + {\rm h.c.}\\
\sum_q \frac{1}{\Lambda^2}
\bar\chi_1\gamma_\mu\chi_2~\bar q\gamma^\mu q + {\rm h.c.}.
\end{eqnarray}
with other choices suppressed by the mass splitting of the dark matter states and/or the small 
typical velocities of WIMPs in the galactic halo. 
As previously discussed, the former is chirality-suppressed while the latter is unsuppressed, and thus 
these lead to very different phenomenologies for annihilations as compared to pair production at colliders. 
We consider a specific representative point from \cite{Frandsen:2013cna}, 
choosing a dark matter mass of 5 GeV and a WIMP-nucleon scattering cross section of 
$3\times10^{-43}$ cm$^2$. The suppression scales needed to induce this scattering rate, as well 
as the pair production cross section at the 14 TeV LHC and the pair annihilation cross section 
corresponding to that suppression scale are presented in table \ref{tab:exo}. We point out that these 
are both equally good fits to the CDMS Si events, and yet the choice of operators leads to 
differences of one order of magnitude in collider signal strength and five orders of magnitude in 
indirect detection, with the effects running in opposite directions. This is a concrete example 
of the effects considered in section \ref{sec:effective} that is of particular current interest.

\begin{table}
\caption{\label{tab:exo}Candidate models for Exothermic Dark Matter in CDMS Si}
\begin{tabular}{|c|c|c|c|}
\hline
~~~Operator~~~&~~~$\Lambda$ (GeV)~~~&
~~$\sigma(pp\to\chi_1\chi_2)$ (fb)~~&~~$\sigma(\chi_1\chi_2\to\bar qq)*v$ (cm$^3$/s)~~\\
\hline
$\bar\chi_1\chi_2~\bar qq$& 260 & 5.3 & $4.88\times10^{-37}$ \\
$\bar\chi_1\gamma_\mu\chi_2~\bar q\gamma^\mu q$& 13570 & 0.44 & $3.9\times10^{-32}$ \\
\hline
\end{tabular}
\end{table}

\section{Conclusions}\label{sec:conclusions}

There is excellent reason to think that we stand at the brink of important discoveries related to dark matter.  
Once these begin, the primary task for particle physics will be to assimilate the message into a successor
theory to the Standard Model which includes dark matter, and then to use that theory to firmly
establish a cosmological picture explaining its relic density.  As we assemble information from direct detection,
indirect detection, and colliders, crossing interactions will play a decisive role.

In this article we have explored several different visions of dark matter, ranging from sketches of interactions
based on effective field theories to UV-complete models such as the MSSM.  In all cases, we could identify
cases where confusion was likely to arise.  But rather than representing serious problems in our ability
to reconstruct the underlying theory of dark matter from a discovery, the examples we provide instead show the necessity and
strength of the multi-pronged search program which is
currently underway.  In the end, the apparent contradictions arising from simple extrapolations
will turn into the tool which will guide us on the path to constructing a theory of dark matter in the context of a new Standard Model for particle physics.

\section*{Acknowledgements}

TMPT is grateful to conversations with Tune Kamae.
WS and SP are partly supported by the US Department of Energy under contract DE-FG02-04ER41268. 
The research of TMPT. is supported in part by NSF
grant PHY-0970171 and by the University of California, Irvine through a Chancellor's fellowship.

\end{document}